\documentclass[preprint,12pt]{elsarticle}

\usepackage{amssymb}

\usepackage{url}
\usepackage{subfigure}

\journal{Future Generation Computing Systems}

\begin{document}

\begin{frontmatter}

\title{Performance and Stability of the Chelonia\\ Storage Cloud}

\author[UiO1,UiO2]{J.~K. Nilsen}
\ead{j.k.nilsen@usit.uio.no}

\author[UU1]{S. Toor}
\ead{salman.toor@it.uu.se}

\author[NIIF]{Zs. Nagy}
\ead{zsombor@niif.hu}

\author[UU2]{B. Mohn}
\ead{bjarte.mohn@fysast.uu.se}

\author[UiO1]{A.~L. Read}
\ead{a.l.read@fys.uio.no}

\address[UiO1]{University of Oslo, Dept. of Physics,  P. O. Box 1048, Blindern, N-0316 Oslo, Norway}
\address[UiO2]{University of Oslo, Center for Information Technology, P. O. Box 1059, Blindern, N-0316 Oslo, Norway}
\address[UU1]{Dept. Information Technology, Div. of Scientific Computing Uppsala University, Box 256, SE-751 05 Uppsala, Sweden}
\address[NIIF]{Institute of National Information and Infrastructure Development NIIF/HUNGARNET, Victor Hugo 18-22, H-1132 Budapest, Hungary}
\address[UU2]{Dept. of Physics and Astronomy, Div. of Nuclear and Particle Physics, Uppsala University, Box 535, SE-751 21 Uppsala, Sweden}

\begin{abstract}
  In this paper we present the Chelonia storage cloud middleware. It
  was designed to fill the requirements gap between those of large,
  sophisticated scientific collaborations which have adopted the grid
  paradigm for their distributed storage needs, and of corporate
  business communities which are gravitating towards the cloud
  paradigm. The similarities to and differences between Chelonia and
  several well-known grid- and cloud-based storage solutions are
  commented. The design of Chelonia has been chosen to optimize high
  reliability and scalability of an integrated system of
  heterogeneous, geographically dispersed storage sites and the
  ability to easily expand the system dynamically. The architecture
  and implementation in term of web-services running inside the
  Advanced Resource Connector Hosting Environment Dameon (ARC HED) are
  described. We present results of tests in both local-area and
  wide-area networks that demonstrate the fault-tolerance, stability
  and scalability of Chelonia.
\end{abstract}


\end{frontmatter}

\section{Introduction}
\label{sec:introduction}
As computationally demanding research areas expand, the need for
storage space for results and intermediate data increases. Currently
running experiments in areas like high energy physics, atmospheric
science and molecular dynamics already generate petabytes of data
every year. The increasing number of international and even global
scientific collaborations also contributes to the growing need to
share (and protect) data efficiently and effortlessly.

While different research groups have different requirements for a
storage system, a set of key characteristics can be identified. The
storage system needs to be {\it reliable} to ensure data integrity. It
needs to be {\it scalable} and capable to dynamically {\it expand} to
future needs, and given that many research groups today use
computational grids to process their data, it is highly favorable if
the storage system is {\it grid-enabled}.

Lately, the concept of storage clouds has gone from being completely
unknown to being the subject of significant attention from end-users
and developers alike~\cite{scientificcloud}. A storage cloud addresses
many of the needs mentioned above, but from a user perspective it also
hides the complexity of the machinery involved in making the system
work. Such a framework requires well-defined roles for the involved
parties (the Service Providers and Infrastructure Providers). It also
requires these groups to have explicit understandings of and
commitments to their roles. Apart from the conceptual agreements,
there is a fundamental need for sustainable technology on which the
framework can built.

In this paper we will present the design and performance of the
Chelonia storage cloud~\cite{Cheloniasite, CheloniaCloud}. With
Chelonia we aim at designing a system which fulfills the requirements
of global research collaborations, but which also meets the
specifications of a storage cloud, e.g., user-centric interfaces and
transparency. With Chelonia it is possible to build anything from
simple storage systems for sharing holiday pictures to large-scale
storage systems for storing petabytes of scientific data.

This paper is organized as follows: First we give a brief introduction
to distributed data storage solutions in Section~\ref{sec:solutions},
before we in Section~\ref{sec:systemoverview} give an architectural
overview of Chelonia and its services. Section~\ref{sec:operation}
exemplifies important features of Chelonia, while
Sections~\ref{sec:systemperformance} and \ref{sec:systemstability}
present performance and stability of Chelonia, respectively. A
comparison with other storage solutions on the market is given in
Section~\ref{sec:relatedwork}. Section~\ref{sec:futurework} presents
ongoing development work, before the conclusions are given in
Section~\ref{sec:conclusion}.

\section{Distributed Data Storage Solutions}
\label{sec:solutions}
Over the years, different concepts have evolved to deal with the
challenge of handling increasing volumes of data and the fact that the
data tend to be generated and accessed over vast geographic
regions. The largest storage solutions nowadays can be roughly divided
between the data grids developed and used by scientific communities,
and cloud storage arising from the needs of corporate business
communities. While both concepts have the same goal of distributing
large amounts of data across distributed storage facilities, their
focuses are slightly different.

Data grid solutions focus on sharing data stored at several large
storage facilities which are usually supported by public funding and
run by different organizations. A grid storage system provides its
clients with access to data stored at remote storage systems. A
traditional architecture (see e.g.~\cite{datagrid,egeeddm}) typically
consists of an {\it indexing service}, indexing files from storage
resources, a {\it file transfer service} for transferring files, a
{\it replication service} for managing replica locations, and a (often
centralized) {\it metadata catalog} imposing a global namespace on top
of the resources. While they enable the sharing of resources between
large number of users, data grids are often considered to be rather
cumbersome in use and maintenance. For example, there is no common
method of establishing a global namespace and this is achieved only
additional effort of the organization that cares for its own data.

Cloud storage focuses more on providing large amounts of storage to
other organizations, and one cloud storage facility is usually run by
a single organization. The main building blocks of a cloud are the
services. The cloud {\it actors} access the services both to add
resources and utilize resources. The services provides
Quality of Service (QoS) guarantees through Service Level Agreements
(SLA's). In clouds, storage is provided through the concept of Data as
a Service (DaaS), which together with Infrastructure as a Service
(IaaS), Hardware as a Service (HaaS) and Software as a Service (SaaS)
can form a Platform as a Service (PaaS). Hence, by combining services,
the service user can create a customized virtual platform. While this
provides more flexibility for the cloud user and service providers
than the grid concept, it limits the ability of sharing
resources. When a user has set up a virtual computing platform, this
platform is typically limited to be used by this user. Data security
is usually realized through isolating the virtual platform to be used
only by the one user.

Storage clouds are an emerging paradigm, and even though the focus
differs from the data grids, the paradigm has learned from the
experience of the grid paradigm and improved on several features like
usability and payment plans. However, arising from the needs of
corporate business the storage clouds lack features like file-sharing
and high-level tools needed by the scientific communities. The
Chelonia storage cloud is designed to combine the best of two
paradigms for a truly distributed, self-healing, flexible storage
solution, with a replicated metadata catalog, easy-to-use storage
resources and an operating system-agnostic implementation.

\section{Architecture of Chelonia}
\label{sec:systemoverview}
Chelonia consists of a set of SOAP-based web services residing within
the Hosting Environment Daemon (HED)~\cite{HEDdesigndoc} service
container from the ARC grid middleware. Together, the services provide
a self-healing, reliable, robust, scalable, resilient and consistent
data storage system. Data is managed in a hierarchical global
namespace with files and subcollections grouped into
collections\footnote{A concept very similar to files and directories
  in most common file systems.}. A dedicated root collection serves as
a reference point for accessing the namespace. The hierarchy can then
be referenced using Logical Names. The global namespace is accessed
the same manner as in local filesystems.

\begin{figure}[htb]
\begin{center}
\includegraphics[width=0.50\columnwidth]{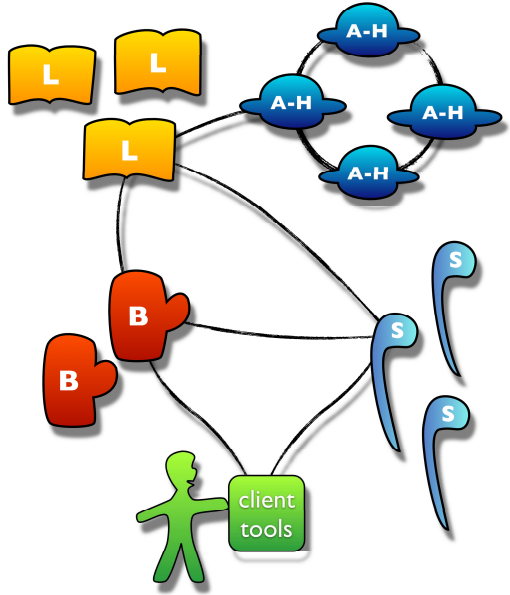}
\end{center}
\caption{Schematic of Chelonia's architecture. The figure shows the main
  services of Chelonia; The Bartender (cup), the Librarian (book), the
  A-Hash (space ship) and the Shepherd (staff). The communication
  channels are depicted by black lines.}
\label{fig:fig1}
\end{figure}

Being based on a service-oriented architecture, the Chelonia storage
cloud consists of a set of services as shown in
Figure~\ref{fig:fig1}. The Bartender (cup) provides the high-level
interface to the user; the Librarian (book) handles the entire storage
namespace, using the A-Hash (space ship) as a meta-database; and the
Shepherd (staff) is the front-end for the physical storage node. Note
that any of the services can be deployed in multiple instances for
high availability and load balancing. Before going into the technical
details of Chelonia itself, it may be beneficial to have a brief look
at the middleware providing the communication layer of Chelonia.

\subsection{The Advanced Resource Connector}
\label{sec:arc}
The Chelonia communication layer is provided by the next generation
Advanced Resource Connector (ARC) Grid middleware, developed by
NorduGrid~\cite{NorduGridsite} and the EU-supported KnowARC
project~\cite{KnowARCsite}. The next generation ARC is based upon
web-services which make frequent use of pluggable components for
offering certain capabilities. The ARC services, including Chelonia
(Section~\ref{sec:coreservices}), run inside a container called the
Hosting Environment Daemon (HED).

There are three basic kinds of plugable components for HED: Data
Management Components (DMC's) are used to transfer the data using
various protocols; Message Chain Components (MCC's) are responsible
for the communication within services as well as between
the clients and the services; and Policy Decision Components (PDC's)
are responsible for the security model within the system. In Chelonia,
the Shepherd uses DMC's to transfer files, all client-service and
inter-service communication goes through the SOAP MCC and the
Bartender uses the PDC to decide if a user has access or not.

\subsection{Core Services}
\label{sec:coreservices}
The main components of Chelonia are the four services, the A-Hash, the
Librarian, the Bartender and the Shepherd. They each have separate
roles based on the distinct characteristics of a distributed storage
system. When compared with traditional data grid solutions, the
Librarian may be viewed as an indexing service and the Shepherd as the
manager of the file transfer service. The replication service is
provided by the Shepherd, Librarian and Bartender services acting
together and the Librarian and A-Hash function as a metadata
catalog. When compared with cloud storage solutions, Chelonia provides
DaaS with the Bartender providing a well-defined API for easy-to-use
access. Acting together, the services provide an easy-to-use,
lightweight storage system without single points of failure.

\subsubsection{The A-Hash}
\label{sec:theahash}
The A-Hash service is a database that stores objects which contain
key-value pairs. In Chelonia, it is used to store the global
namespace, all file metadata and information about itself and storage
elements. Being such a central part of the storage system, the A-Hash
needs to be consistent and fault-tolerant. The A-Hash is replicated
using the Oracle Berkeley DB~\cite{berkeley} (BDB), an open source
database library wih a replication API\@. The replication is based on
a single master, multiple clients framework where all clients can read
from the database and only the master can write to the database. While
a single master ensures that the database is consistent at all times,
it raises the problem of having the master as a single point of
failure. If the master is unavailable, the database cannot be updated,
files and entries cannot be added to Chelonia and file replication
will stop working. The possibility of the master failing cannot be
completely avoided, so to ensure high availability means must be taken
to find a new master if the first master becomes unavailable. BDB uses
a variant of the Paxos algorithm~\cite{Paxos} to elect a master
amongst peer clients: Every database update is assigned an increasing
number. In the event of a master going offline, the clients sends a
request for election, and a new master is elected amongst the clients
with the highest numbered database update.

\subsubsection{The Librarian}
\label{sec:librarian}
The Librarian service manages the hierarchy and metadata of files and
collections, handles the Logical Names and monitors the Shepherd
services. The Librarian service is stateless, instead it stores all
the persistent information in the A-Hash. This makes it possible to
deploy any number of independent Librarian services to provide
high-availability and load-balancing. In this case all the Librarians
should communicate with the same set of A-Hashes in order to use the
same database of metadata. As only the master A-Hash can be written to
and the Librarian cannot know {\it a priori} which A-Hash replica is
the master, the Librarian needs to get this information from one of
the A-Hashes. For this reason, the master A-Hash stores the list of
all available A-Hashes, so that the information is
replicated to all A-Hash replicas. As all A-Hash replicas are
readable, the Librarian only needs to know about a few of the A-Hashes
at start-up to be able to get this list. During run-time the Librarian
holds a local copy of the A-Hash list and refreshes it both regularly
and in the case of a failing connection.

\subsubsection{The Shepherd}
\label{sec:shepherd}
Each instance of the Shepherd service manages a particular storage
node and provides a uniform interface for storing and accessing file
replicas. On a storage node there must be at least one independent
storage element service (with an interface such as HTTP(S), ByteIO,
etc.)\ which performs the actual file transfer. A storage node then
consists of a Shepherd service and a storage element service connected
together. Storage element services can either be provided by ARC or by
third-party services. For each kind of storage element service, a
Shepherd backend module is needed to enable the Shepherd service to
communicate with the storage element service, e.g., to initiate file
uploads, downloads and removal, and to detect whether a file transfer
was successful or not. Currently there are three Shepherd backends:
One for the ARC native HTTP(S) server called Hopi; one for the Apache
web server; and one for a service which implements a subset of the
ByteIO interface. In addition to storing files and providing access to
them, the Shepherd is responsible for checking if a file replica is
valid and, if necessary, initiating replication of the file to other
Shepherds.

\subsubsection{The Bartender}
\label{sec:bartender}
The Bartender service provides a high-level interface of the storage
system for the clients (other services or users). The clients can
create and remove collections (directories), create, get and remove
files, and move files and collections within the namespace using
Logical Names. Access policies associated with files and collections
are evaluated by the Bartender (using the PDC plugin of HED) every
time a user wants to access them. The Bartender communicates with the
Librarian and Shepherd services to execute the client's requests.  The
file content itself does not go through the Bartender; file transfers
are directly performed between the storage nodes and the clients. 

The Bartender also supports so-called gateway modules which make it
possible to communicate with third-party storage solutions, thus
enabling the user to access multiple storage systems through a single
Bartender client. These modules are protocol-oriented in the sense
that external storage managers which support a certain protocol will
be handled using the gateway module based on that protocol. While
excluding some of the features provided by accessing storage managers
directly, this approach reduces the number of gateway modules
required for different storage managers. The currently available
gateway module is based on the GridFTP protocol\cite{gridftp}.

\subsection{Security}
\label{sec:security}
As is the case for all openly accessible web services, the security
model is of crucial importance for the Chelonia storage cloud. While
the security of the communication with and within the storage system
is realized through HTTPS with standard X.509 authentication, the
authorization related security architecture of the storage can be
split into three parts; the inter-service authorization; the
transfer-level authorization; and the high-level authorization:

\begin{itemize}
\item The inter-service authorization maintains the integrity of the
  internal communication between services. There are several
  communication paths between the services in the storage system. The
  Bartenders send requests to the Librarians and the Shepherds, the
  Shepherds communicate with the Librarians and the Librarians with
  the A-Hash. If any of these services are compromised or a new rogue
  service gets inserted in the system, the security of the entire
  system is compromised. To enable trust between the services, they
  need to know each other's Distinguished Names (DN's). This way a
  rogue service would need to obtain a certificate with that exact DN
  from some trusted Certificate Authority (CA).

\item The transfer-level authorization handles the authorization in
  the cases of uploading and downloading files. When a transfer is
  requested, the Shepherd will provide a one-time Transfer URL (TURL)
  to which the client can connect. In the current architecture, this
  TURL is world-accessible. This may not seem very secure at
  first. However, provided that the TURL has a very long, unguessable
  name, that it is transfered to the user in a secure way and that it
  can only be accessed once before it is deleted, the chance of being
  compromised is fairly low.

\item The high-level authorization considers the access policies for
  the files and collections in the system. These policies are stored
  in the A-Hash, in the metadata of the corresponding file or collection,
  providing a fine-grained security in the system.
\end{itemize}

\subsection{Accessing Chelonia}
\label{sec:clienttools}
Being the only part a user will (and should) see from a storage
system, the client tools are an important part of the Chelonia storage
cloud. In addition to a vanilla command-line interface, two ways
of accessing Chelonia are supported.

\subsubsection{FUSE Module}
\begin{figure}[htb]
\begin{center}
\includegraphics[width=0.70\columnwidth]{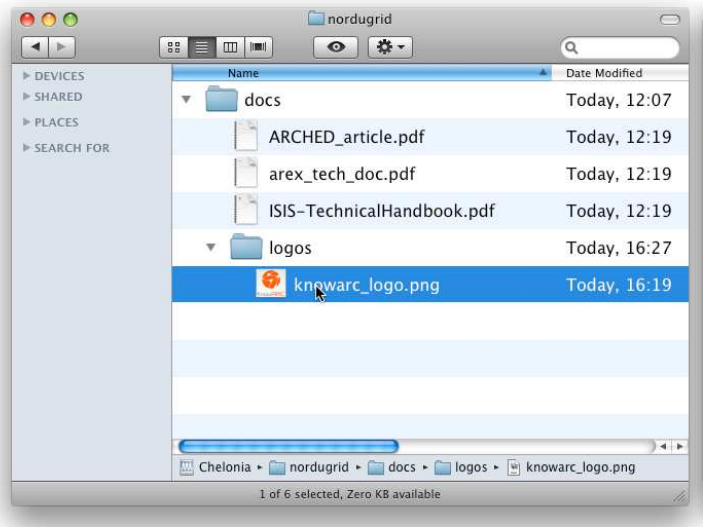}
\end{center}
\caption{Screenshot of the Chelonia FUSE module in use. Through the
  FUSE module Chelonia offers users a drag and drop functionality to
  upload or download files to the storage cloud.}
\label{fig:fusescreenshot}
\end{figure}

The FUSE module provides a high-level access to the storage
system. Filesystem in Userspace (FUSE)~\cite{FUSE} provides a simple
library and a kernel-userspace interface. Using FUSE and the ARC
Python interface, the FUSE module allows users to mount the storage
namespace into the local namespace, enabling the use of graphical file
browsers as shown in the screenshot in
Figure~\ref{fig:fusescreenshot}.

\subsubsection{Grid Job Access}
To access data through the ARC middleware client tools, one needs to
go through Data Manager Components (DMC's). These are
protocol-specific plugins to the client tools. For example, to access
data from a HTTPS service, the HTTP DMC will be used with a URL
starting with \verb!https://!, to access data from an SRM service, the
SRM DMC will be used with a URL starting with
\verb!srm://!. Similarly, to access Chelonia, the ARC DMC will be used
with a URL starting with \verb!arc://!.

The ARC DMC allows grid jobs to access Chelonia directly. As long as
A-REX, the job execution service of ARC, and ARC DMC are installed on
a site, files can be both downloaded and uploaded by specifying the
corresponding URL's in the job description. In this case, the
Bartender URL needs to be embedded in the URL as a URL option. For
example, if a job requires an input file \verb!/user/me/input.dat!,
the URL specified in the job description will be as follows (given
that the file can be found by a Bartender with URL
\verb!https://storage/Bartender!):
\begin{verbatim}
arc:///user/me/input.dat?BartenderURL=https://storage/Bartender
\end{verbatim}
 
\section{Chelonia in Operation}
\label{sec:operation}
Both the Chelonia storage system and its clients can be installed from
binary packages (available for several different platforms) or after
compiling the source packages. A fully operational storage cloud
requires a minimum installation of one instance of every service
described above. The Chelonia Administrator
manual~\cite{cheloniaadmin} gives detailed instructions on how
to install, configure and run the services. In order for users to
interact with Chelonia, several user tools are provided. These are
documented both in the Chelonia user
manual~\cite{cheloniausermanual} and Linux man pages and directly through
command line calls.

For the user, transfering files to and from Chelonia are simple
operations. For example, if a user wants to upload a file
\verb!orange.jpg! to Chelonia, he/she can use, e.g., the Chelonia
CLI. Assuming that the URL of one or more Bartenders and the required
number of file replicas are written in a configuration file, the user
gives the command
\begin{verbatim}
chelonia put orange.jpg /user/me/orange.jpg
\end{verbatim}
Note that there is
no need for the user to know where files are physically stored or will
be stored in Chelonia.

Under the hood of Chelonia, the Bartender receiving the request from
the CLI contacts a Librarian to create an entry in the Chelonia
namespace. If the Librarian confirms the new entry, the Bartender then
contacts a Shepherd to get a transfer URL which is returned to the
CLI. When the CLI has uploaded the file, the Shepherd queries the
Librarian to find out how many replicas are needed and, if needed,
initiates a file transfer to another Shepherd. In the case of
downloading a file, the Bartender gets the locations of the file from
a Librarian, chooses one of them and contacts the corresponding
Shepherd for a transfer URL.

When in operation, the Chelonia storage cloud is a pulsing system where
heartbeats are periodically sent from each Shepherd to a Librarian
together with information about replicas whose state changed since the
last heartbeat. Heartbeats are stored in the A-Hash, thus making them
visible to all Librarians in the system. If any of the Librarians
notices that a Shepherd is late with its heartbeat, it will mark all
the replicas in that Shepherd as offline.

In addition to the heartbeat, the Shepherds periodically check with
the Librarians to see if there are sufficient replicas of the files in
Chelonia and if the checksums of the replicas are correct. If a file
is found to have too few replicas, the Shepherd informs a Bartender
about this situation and together they ensure that a new replica is
created at a different storage node. A file having too many replicas
will also be automatically corrected by Chelonia - the first Shepherd
to notice this will mark its replica(s) as unneeded and later delete
it (them). Replicas with invalid checksums are marked as invalid, and
as soon as possible replaced with a valid replica.

With the Chelonia gateway module, a user can mount external storage
systems into the Chelonia namespace. For example, if a Chelonia user has
access to a set of files stored in dCache~\cite{dCache} (see
Section~\ref{sec:relatedwork}) under \verb!/fruits! he can add a mount
point (say \verb!/my/dCache!) to easily access these data
through the Chelonia namespace, i.e., using standard Chelonia commands
like
\begin{verbatim}
chelonia get /my/dCache/fruits/apple.jpg
\end{verbatim}
More technically, when the Bartender looks up the entry
\verb!/my/dCache! which is a mount point to dCache, it will use the
corresponding gateway module to generate an external URL which the
client tool will use to contact dCache directly. This way, Chelonia
can include third-party storage namespaces in its global namespace by
simply storing a single entry.

\section{System Performance}
\label{sec:systemperformance}

\subsection{Adding and Querying the Status of Files}
\label{sec:addingfiles}
In a hierarchical file system files are stored in levels of
collections and sub-collections. The time to add or get a file depends
mainly on two factors; the number of entries in the collection where
the file is inserted, and the number of parent collections to the
collection where the file is inserted (the depth of the
collection). Based on these two factors we have run two different
tests:

\begin{itemize}
\item {\bf Depth test} tests the performance when creating many levels
  of sub-collections. The test adds a number of sub-collections to a
  collection, measures the time to add and stat the sub-collections,
  then adds a number of sub-sub-collections to one of the
  sub-collections and so forth. To query a collection at a given level
  means that all the collections at the lower levels needs to be
  queried first. In Chelonia, each query causes a message to be sent
  through TCP\@. Hence, it is expected that time will increase
  linearly as the level of collections increases. As every message is
  of equal size, this test ideally depends only on network latency.
\item {\bf Width test} tests the performance when adding many entries
  (collections) to one collection. The test is carried out by adding a
  given number of entries to a collection and measuring the time to add
  each entry and the time to stat the created entry. When adding an
  entry to a collection, the system needs to check first if the entry
  exists. In Chelonia, this means that the list of entries in the
  collection needs to be transferred through TCP\@. It is therefore
  expected that the time to add an entry will increase linearly as
  this list increases and ideally the time will depend only on the
  bandwidth of the network.
\end{itemize}

Both tests were run in two types of environments: In the Local Area
Network (LAN) setup four computers were connected to the same switch.
A centralized A-Hash service, a Librarian, a Bartender, and the client
were each run on a separate computer. In the Wide Area Network (WAN)
setup, the client and Bartender were located in Uppsala, Sweden, while
the Librarian and a replicated A-Hash consisting of three replicas were
located in Oslo, Norway.

\begin{figure}[ht]
\centering
\subfigure[Services running on LAN]{
\includegraphics[width=0.47\columnwidth]{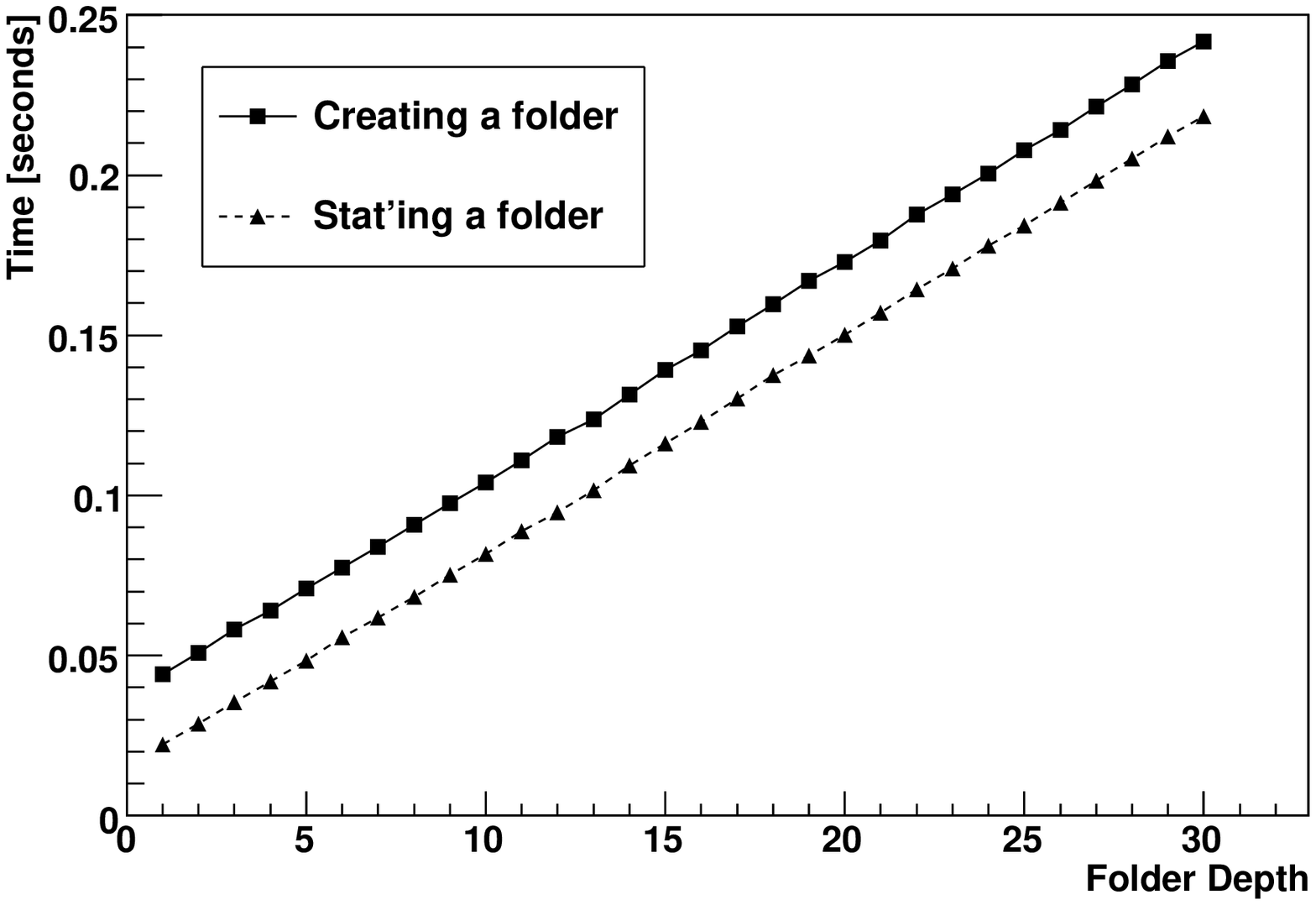}
\label{fig:depthTestLan}
}
\subfigure[Services running on WAN]{
\includegraphics[width=0.47\columnwidth]{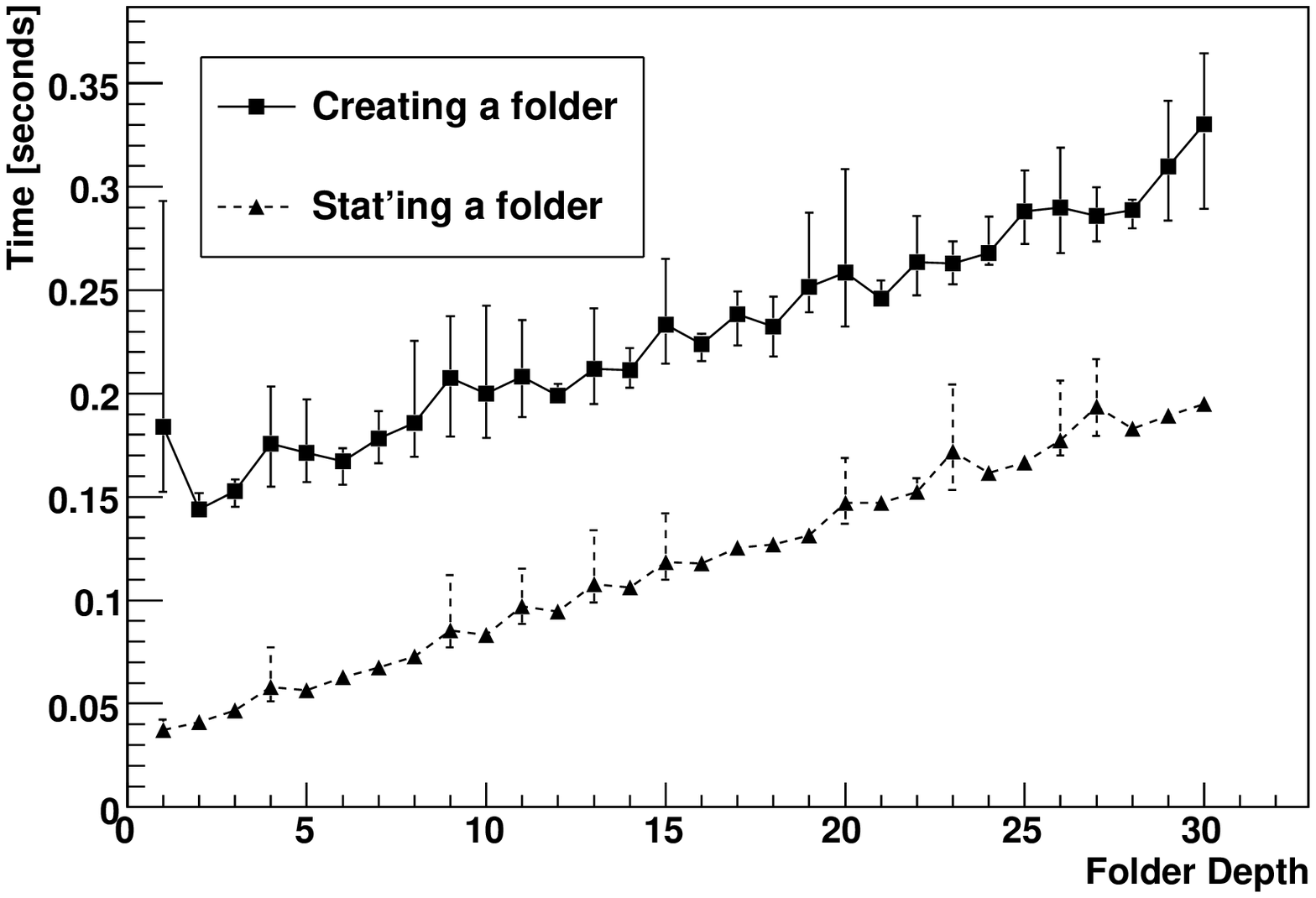}
\label{fig:depthTestWan}
}
\caption{Time to add an entry to a collection (continuous line) and
                time to get status of a collection (dashed line) given the
                hierarchical depth of the collection. }
\end{figure}

The test result for the depth test is shown in Figures
\ref{fig:depthTestLan} (LAN) and \ref{fig:depthTestWan} (WAN)\@. The
continuous lines show the time to create an entry, while the dashed
lines show the time to get the status of the collection. All the plots
are averages of five samples, with the error bars representing the
minimum and maximum values. The LAN test shows a near-perfect linear
behaviour, with the error bars too small to be seen. As mentioned
earlier, since the packet size for each message is constant in this
test, the main bottleneck (apart from Chelonia itself) is the network
latency. Since all computers in the LAN test are connected to the same
switch we can assume that the latency is near constant. Hence, the LAN
test shows that in a very simple network scenario Chelonia works as
expected, with the network being the major bottleneck. Notice also
that creating an entry consistently takes 0.021~s longer than
getting the status of the collection, again corresponding to the extra
message needed to create an entry.

In the WAN test, the complexity is a bit increased, as in addition to
sending messages over WAN, the A-Hash is now replicated. The time
still increases linearily, albeit with more fluctuation due to the WAN
environment. Creating an entry at the first level in the hierarchy now
takes 0.11~s longer than getting the status of the collection,
corresponding to the fact that the entry needs to be replicated three
times. However, at higher levels, getting the status over WAN is
actually faster than getting the status over LAN. The effect of the
replicated A-Hash will be discussed in more detail in Section
\ref{sec:ahashtest}.

\begin{figure}[ht]
\centering
\subfigure[Services running on LAN]{
\includegraphics[width=0.47\columnwidth]{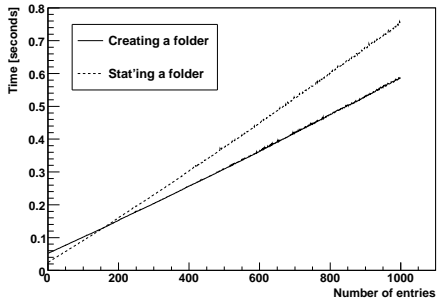}
\label{fig:widthTestLan}
}
\subfigure[Services running on WAN]{
\includegraphics[width=0.47\columnwidth]{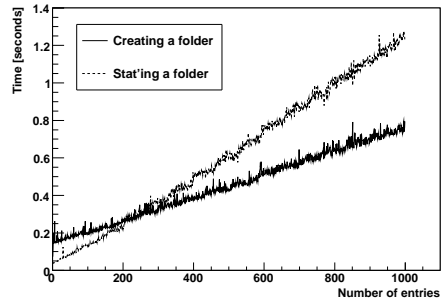}
\label{fig:widthTestWan}
}
\caption{Time to add an entry to a collection (continuous line) and
                time to get status of a collection (dashed line) given the
                number of entries already in the collection.}
\end{figure}

The test results for the width test are shown in Figures
\ref{fig:widthTestLan} (LAN) and \ref{fig:widthTestWan} (WAN)\@. The
continous lines show the time to add entries to a collection and the
dotted lines show the time to get the status of a collection
containing the given number of entries. The operations were repeated
five times, with the plots showing the averages of these five
samples. As expected, the time increase linearily with increasing
number of entries. The results of the WAN test fluctuate more than
those of the LAN test, which is to be expected; in the LAN test all
services are on computers connected to the same switch, while in the
WAN test, the services are distributed between two different
countries. However, the WAN test shows similar linearity, albeit with
a slightly higher response time. It is worth noticing that for the
width test, in contrast to the depth test, we see no benefit of using
a replicated A-Hash. This is due to the fact that the bandwith is the
limiting factor in this test.

An interesting feature of both tests is that while creating an entry
takes more time when there are only few entries in the collection,
creating new entries is actually faster than stating the collection
when the collection has many entries. This is due to the fact that
getting the status of a collection requires the metadata of the
collection (and hence the list of entries in the collection) to be
transfered first from the A-Hash to the Librarian, and second from the
Librarian to the Bartender and last from the Bartender to the
client. When creating an entry, neither the Bartender nor the client
needs this list of entries, so that less data is transfered between
services. However, for fewer entries, creating new entries is more
expensive since the Bartender needs to query the Librarian twice,
first to check if it is allowed to add the entry and second to
actually add it.

\subsection{File Replication}
\label{sec:filereplication}
The concept of automatic file replication in Chelonia was presented in
Section~\ref{sec:operation}. In this section we will demonstrate both
how Chelonia works with different file states to ensure that a file
always has the requested number of valid (ok) replicas and how
Chelonia distributes the replicas in the system in order to ensure
maximum fault tolerance of the system.

The test system consists of one Bartender, one Librarian, two A-Hashes
(one client and one master) and five Shepherds. All services are
deployed within the same LAN. As a a starting point 10 files of 114~MB
are uploaded to the system and for each file 4 replicas are
requested. Thus the initial setup of the test system contains 40
replicas with an initial distribution of file replicas as shown in
Table~\ref{tab:initialfiledistribution}.

\begin{table}[ht]
\centering
\begin{tabular}{c r  r }
\hline
Shepherd & Initial  & Final\\
\hline
\hline
S1 & 9  & 9\\

S2 & 8  & 7\\

S3 & 8  & 8\\

S4 & 7 & 9\\

S5 & 8 & 7\\
\hline
Total & 40 & 40\\
\end{tabular}
\caption{Initial and final load distribution of 40 files on 5 Shepherds}
\label{tab:initialfiledistribution}
\end{table}

The first phase of the file replication test was to kill one of the
Shepherd services, S3, of the test system. Chelonia soon recognized
the loss of this service (no heartbeat received within one cycle) and
started compensating for the lost replicas. File replicas in Chelonia
have states ALIVE, OFFLINE, THIRDWHEEL or CREATING which are recorded
in the A-Hash. Initially the test system had 40 ALIVE replicas (10
files with 4 replicas each), but when the Librarian did not get the S3
heartbeat it changed the state of the 8 replicas stored in S3 to
OFFLINE.

At this point a number of files stored in our Chelonia setup had too
few ALIVE replicas. As explained above, the Shepherds check
periodically that files with replicas stored on its storage element
have the correct number of ALIVE replicas. Thus, in the next cycle the
S1, S2, S4 and S5 Shepherds started to create new file replicas which
initially appeared in the system with the state CREATING.

\begin{figure}[htb]
\begin{center}
\includegraphics[width=0.8\columnwidth]{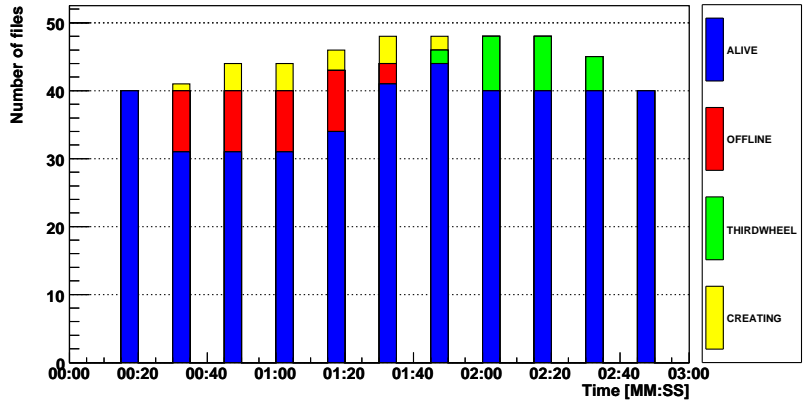}
\end{center}
\caption{Number of replicas and their corresponding states when
  querying the test system every 15~seconds. Shepherd S3 is turned
  offline between 00:15 and 00:30 and back online between 01:15 and
  01:30.}
\label{fig:FileReplication}
\end{figure}

Figure~\ref{fig:FileReplication} gives a graphical overview of the
number of replicas and the replica states in the test system every 15
seconds. At 15~s (00:15) all 40 file replicas were ALIVE as explained
above, but soon thereafter S3 was turned off and the other Shepherds
started creating new replicas. Queried at 30~s (00:30) the system
contained 32 ALIVE, 8 OFFLINE and 1 CREATING file replica.

While the system worked on compensating for the loss of S3, the second
phase of the replication test was initiated by turning the S3 shepherd
online again. The reappearence of the S3 Shepherd can be seen in
Figure~\ref{fig:FileReplication} at 90s (01:30) as a significant
increase in the number of ALIVE replicas. In fact there were now too
many ALIVE replicas in the system and at 105~s (01:45) 2 replicas were
marked as THIRDWHEEL (the Chelonia state for redundant
replicas). THIRDWHEEL replicas are removed from Chelonia as soon as
possible, and during the next 45~s the system removed all such
replicas. A query of the system at 165~s (02:45) shows that the system
once again contained 40 ALIVE replicas. The final distribution of
replicas between Shepherds is given in
Table~\ref{tab:initialfiledistribution}.

\subsection{Multi-User Performance}
\label{sec:multiuser}
While any distributed storage solution must be robust in terms of
multi-user performance, it is particularly important for a
grid-enabled storage cloud like Chelonia where hundreds of grid jobs
and interactive users are likely to interact with the storage system
in parallel. To analyze the performance of Chelonia in such
environments we have studied the response time of the system while
increasing the number of simultaneous users (multi-client).

\begin{figure}[ht]
\centering
\includegraphics[width=0.8\columnwidth]{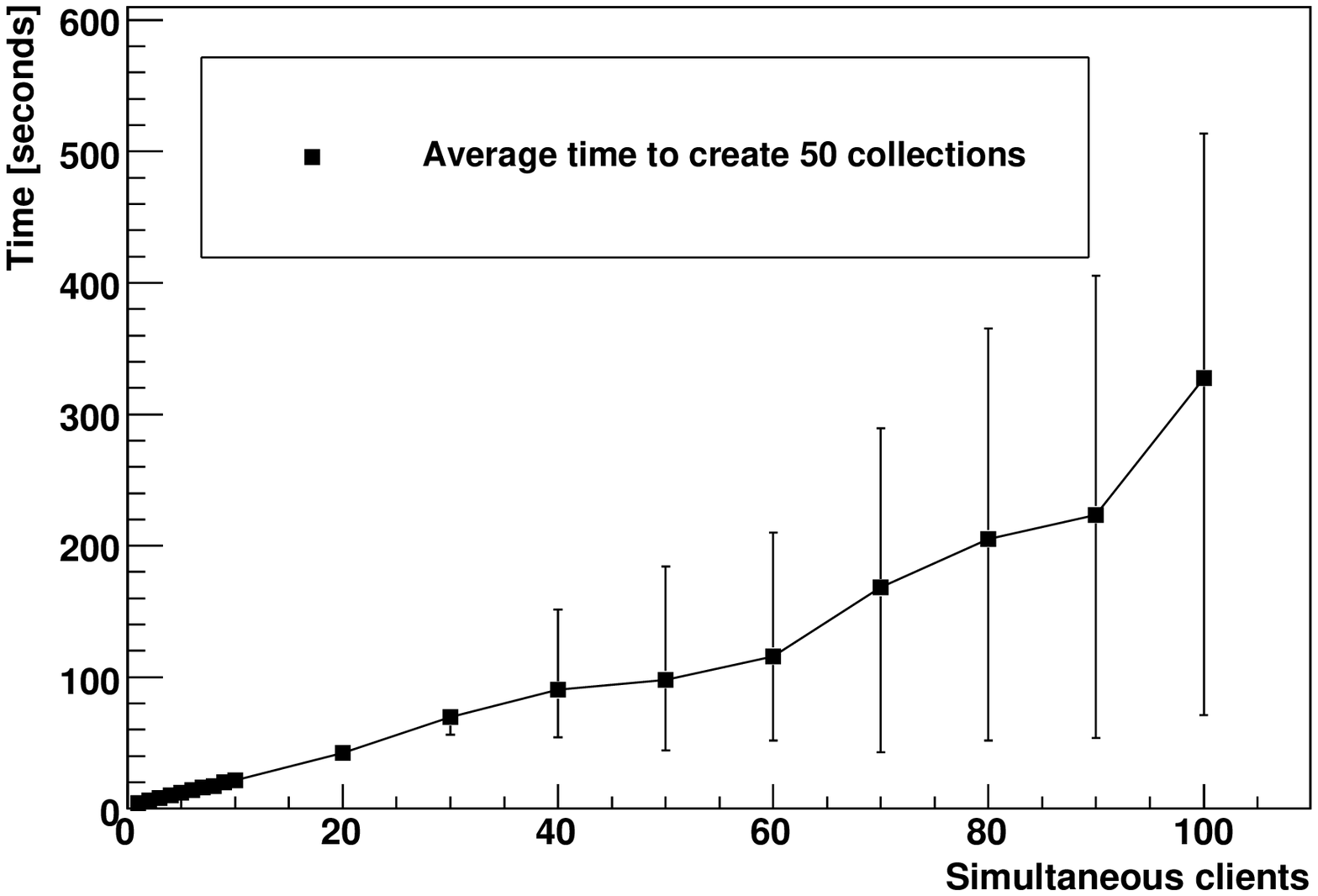}
\label{fig:MultiClientLan}
\caption{Average (square), minimum (lower bar) and maximum (bar)
  system response time as a function of the number of simultaneous
  clients of the system. Each client creates 50 collections
  sequentially.}
\end{figure}

Due to limited hardware resources, multiple clients for the tests were
simulated by running multiple threads from three different
computers. Each client thread creates 50 collections sequentially and
tests were done for an increasing number of simultaneous clients. For each
test the minimum, average and maximum time used by the client was
recorded. Figure~\ref{fig:MultiClientLan} shows the system response
times for up to 100 simultaneous clients using the above-mentioned
testbed deployment. The shown test was run in a LAN environment with
one centralized A-Hash, one Librarian and one Bartender.

The test results show that the response time of the system increases
linearly with an increasing number of simultaneous clients. From 40
clients and onwards the difference between the fastest client and
slowest client starts to become sizeable. When running 50 clients or
more in parallel, it was occasionally observed that a client's request
failed due to the heavy load of the system. When this happened, the
request was retried until successfully completed, as shown by the
slightly fluctuating slope of the mean curve. The same linearity was
seen in the corresponding WAN test (not shown), albeit with a factor
two higher average time, consistent with the results observed in the
depth and width tests in Section~\ref{sec:addingfiles}.

As can be noted in Figure~\ref{fig:MultiClientLan}, for more than 30
clients, the maximum times increase approximately linearly while the
minimum times are close to constant. The reason for this is a
limitation on the number of concurrent threads in the Hosting
Environment Daemon (HED). If the number of concurrent requests to HED
reaches the threshold limit, the requests are queued so that only a given
number of requests are processed at the same time. In the test each
client used only one connection for creating all 50 collections. Hence,
the fastest request was one that had not been queued so that when
the number of requests was above the threshold, the minimum timing did
not depend on the total number of clients. On the other hand, the
slowest request was queued behind several other requests so
that the maximum time increased with the number of simultaneous
clients.

\subsection{Centralized and Replicated A-Hash}
\label{sec:ahashtest}

As the A-Hash stores all metadata about files, file locations and
shepherds, it is important that the A-Hash is fault tolerant and able
to survive even fatal hardware failures. While in theory replicating
the A-Hash provides these features, the replication adds complexity to
the A-Hash in that all data need to be replicated to all A-Hash
instances. Additionally, in the event of a failing A-Hash instance, the
Librarians need to seamlessly find and connect to other A-Hashes.

To test the fault tolerance and performance overhead of the replicated
A-Hash in a controlled environment, four tests have been set up
with services and a client on different computers in the same LAN:
\begin{enumerate}
\item {\bf Centralized}: One client contacting a centralized A-Hash
  was set up as a benchmark, as this is the simplest possible
  scenario.
\item {\bf Replicated, stable}: One client contacting three A-Hash
  instances (one A-Hash master, two A-Hash clients) randomly. All the
  A-Hashes were running during the entire test.
\item {\bf Replicated, unstable clients}: Same setup as in point 2,
  but with a random A-Hash client restarted every 60 seconds.
\item {\bf Replicated, unstable master}: Same setup as in point 2,
  but with the master A-Hash restarted every 60 seconds.
\end{enumerate}
While setups 1 and 2 test the differences in having a centralized
A-Hash and a replicated A-Hash, setups 3 and 4 tests how the system
responds to an unstable environment. In all four setups the system
has services available for reading at all times. However, in setup 3
one may need to reestablish connection with an A-Hash client and in
setup 4 the system is not available for writing during the election of
a new master. During the test the
client computer constantly and repeatedly contacted the A-Hash for
either writing or reading for 10 minutes. During write tests the client
computer reads the newly written entry to ensure it is correctly
written.

\begin{table}[ht]
\centering\small
\begin{tabular}{l |c | c | c}
\hline
\multicolumn{4}{c}{\bf Reading}\\
\hline
 & {\it Minimum (s)} & {\it Average (s)} &
{\it Maximum (s)}\\
\hline

Centralized                  & 0.003399 & 0.003780 & 0.013441 \\

Replicated, stable           & 0.003453 & 0.003738 & 0.013261\\

Replicated, unstable clients & 0.003412 & 0.003754 & 0.289535\\

Replicated, unstable master  & 0.003402 & 0.003763 & 1.971131\\
\hline
\multicolumn{4}{c}{}\\\hline
\multicolumn{4}{c}{\bf Writing}\\
\hline
 & {\it Minimum (s)}  & {\it Average (s)} & {\it Maximum (s)}\\
\hline

Centralized                  & 0.003828 & 0.004260 & 0.014459 \\

Replicated, stable           & 0.016866 & 0.033902 & 1.057602 \\

Replicated, unstable clients & 0.016434 & 0.034239 & 1.131142 \\

Replicated, unstable master  & 0.016293 & 0.044868 & 60.902862\\
\hline
\end{tabular}
\caption{Timings for reading from and writing to a centralized A-Hash
  compared with a stable replicated A-Hash, a replicated A-Hash where
  clients are restarted and a replicated A-Hash where the master is
  restarted.}
\label{tab:ahashreadwright}
\end{table}

Table~\ref{tab:ahashreadwright} shows timings of reading from and
writing to the A-Hash for the four setups. As can be seen, for
reading, all four setups have approximately the same
performance. Somewhat surprisingly, the replicated setups actually
perform better than the centralized setup, even though only one client
computer was used for reading. While reading from more A-Hash
instances is an advantage for load balancing in a multi-client
scenario, one client computer can only read from one A-Hash instance
at a time. Thus, the only difference between the centralized and
replicated setups in terms of reading is how entries are looked up
internally in each A-Hash instance, where the centralized A-Hash uses
a simple Python dictionary, while the replicated A-Hash uses the
Berkeley DB which has more advanced handling of cache and memory.

Looking at the write performance in Table~\ref{tab:ahashreadwright},
there is a more notable difference between the centralized and
the replicated setups. On average, writing to the replicated A-Hash
takes almost 10 times as long as writing to the centralized
A-Hash. The reason for this is that the master A-Hash will not
acknowledge that the entry is written until all the A-Hash instances
have confirmed that they have written the entry. While this is rather
time-consuming, it is more important that the A-Hash is consistent
than fast. It is however worth noticing when implementing services
using the A-Hash, that reading from a replicated A-Hash is much faster
than writing to it.

\section{System Stability}
\label{sec:systemstability}

While optimal system performance may be good for the day-to-day user
experience, the long-term stability of the storage system is an
absolute requirement. It does not help to have a response time of a
few milliseconds under optimal conditions if the services need to be
frequently restarted due to memory leaks or if the servers become
unresponsive due to heavy load. 

To test the system stability, a Chelonia deployment was run
continuously over a week's time. During the test a client regularly
interacted with the system, uploading and deleting files and listing
collections. The deployment consisted of one Bartender, one Librarian,
three A-Hashes and two storage nodes, each consisting of a Shepherd
and a Hopi service. All the services and the client ran on separate
servers in a LAN environment.

\begin{figure}
\includegraphics[angle=90,width=0.9\columnwidth]{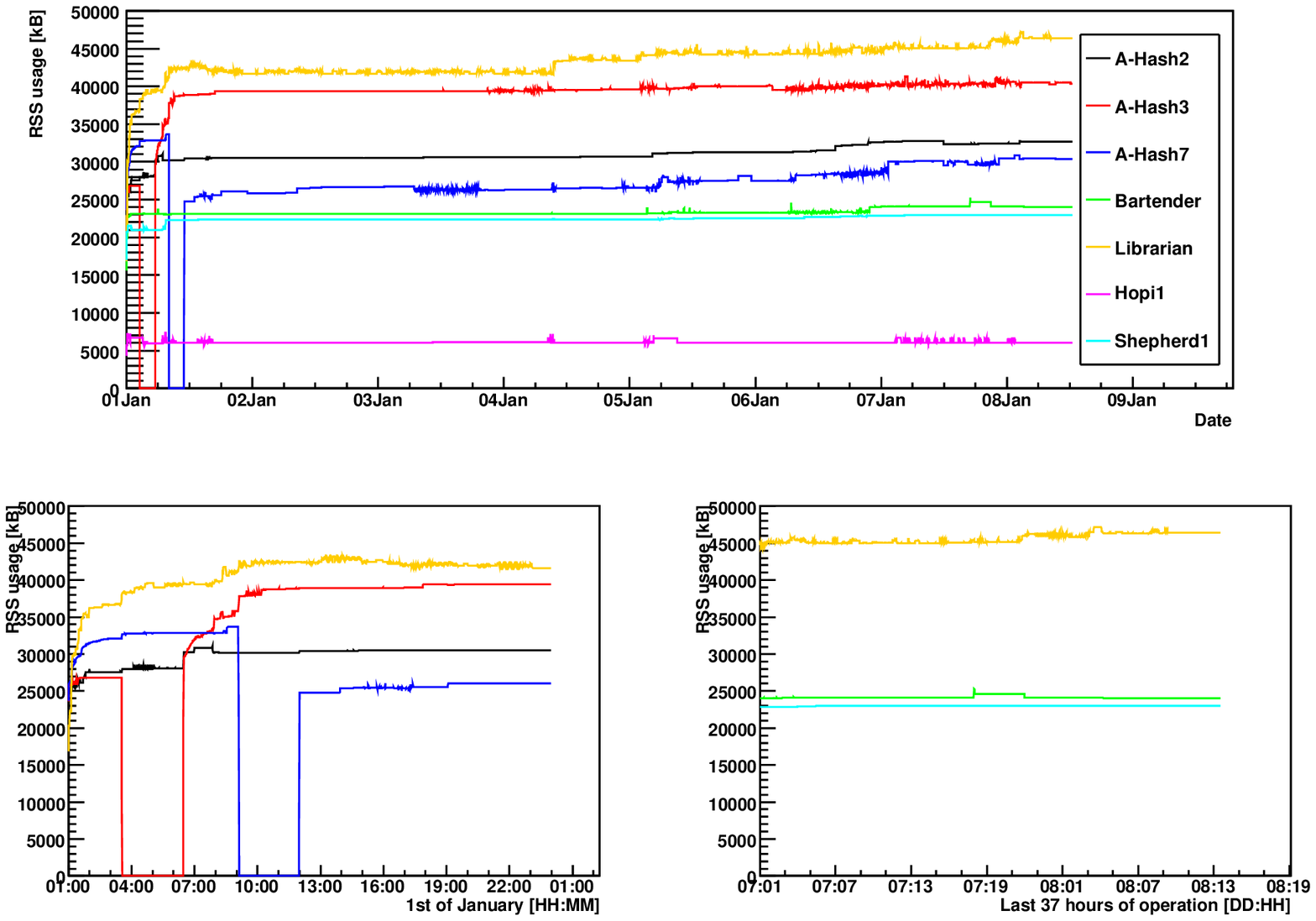}
\caption{Resident memory utilization of the Chelonia services during
  an 8 day run.}
\label{fig:RSSconsumption}
\end{figure}

Figure~\ref{fig:RSSconsumption} shows the overall memory utilization
for seven of the services for the entire run time (top), the A-Hashes and
the Librarian in the first 24 hours (bottom left) and the Librarian,
the Bartender and one of the Shepherds in the last 37 hours (bottom
right). The memory usage was measured by reading the memory usage of
each service process in 5 seconds intervals using the Linux \verb!ps!
command.

The most crucial part of Chelonia, when it comes to handling server
failures, is the replicated A-Hash. If the A-Hash becomes unavailable,
the entire system is unavailable. If a client A-Hash goes down, the
Librarians may need to find a new client. If the master A-Hash goes
down, the entire system will be unavailable until a new master is
elected. The bottom left of Figure~\ref{fig:RSSconsumption} shows the
memory consumption of the three A-Hashes, A-Hash2, A-Hash3 and
A-Hash7, and the Librarian during the first 24 hours of the stability
test. When the test started A-Hash3 (red line) was elected as A-Hash
master, and the Librarian started using A-Hash7 (blue line) for read
operations. After 2.5 hours, A-Hash3 was stopped (seen by the sudden
drop of the red line), thus forcing the two remaining A-Hashes to
elect a new master between them. While not visible on the figure, the
A-Hash, and hence Chelonia, was unavailable for a 10 seconds period
during the election, which incidentally was won by A-Hash2 (black
line). After three additional hours, A-Hash3 was restarted, thus
causing an increase of memory usage for the master A-Hash as it needed
to update A-Hash3 with the latest changes in the database. Eight hours
into the test, the same restart procedure was carried out on A-Hash7
which was connected to the Librarian. This time there was no
noticeable change in performance. However A-Hash3 increased memory
usage when the Librarian connected to it.

\begin{figure}
\includegraphics[width=1.1\columnwidth]{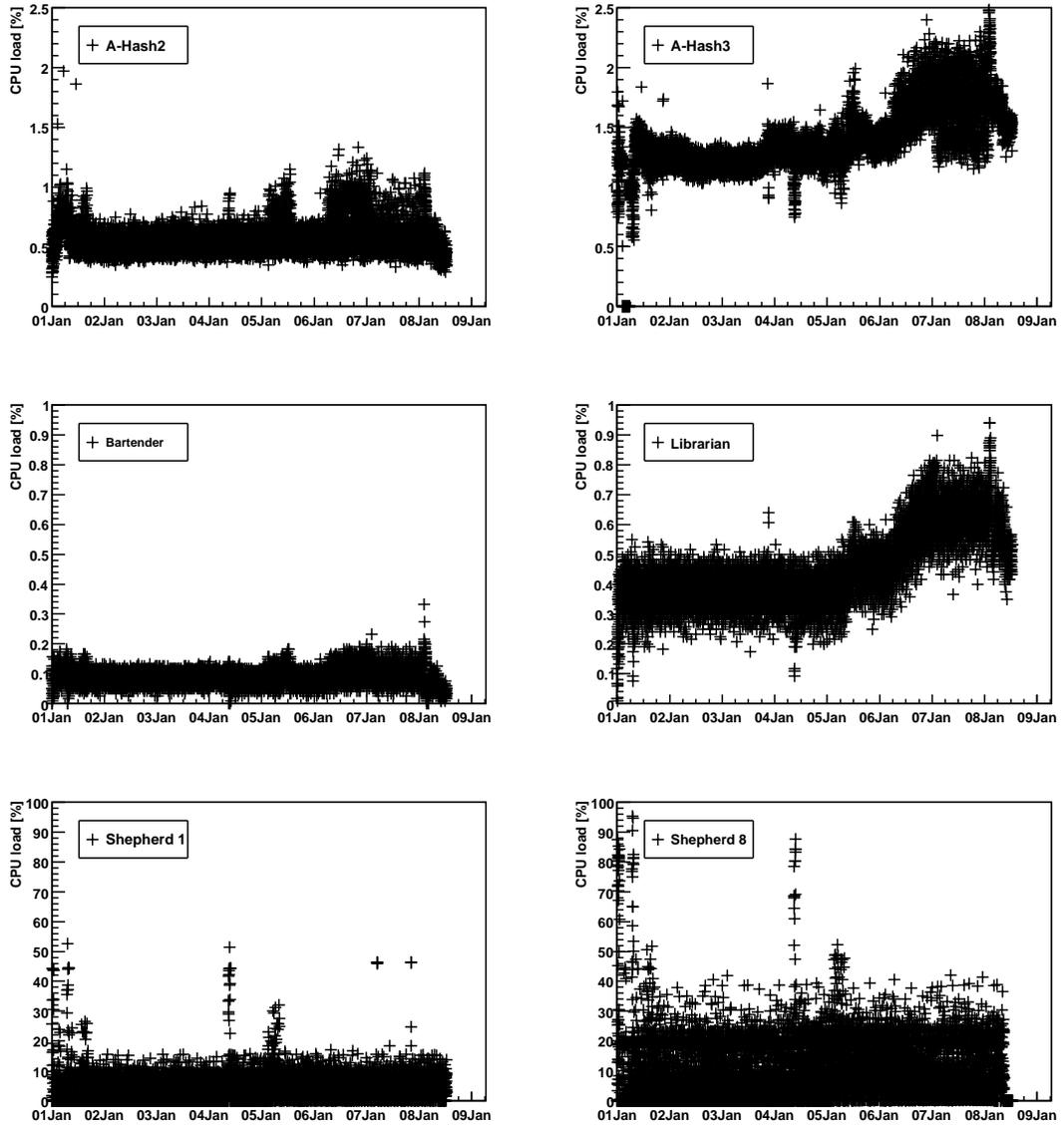}
\caption{ CPU load of the Chelonia services during an 8 day run. Each
  point is an average of the CPU usage of the previous 60 seconds.}
\label{fig:CPUconsumption}
\end{figure}

The bottom right of Figure~\ref{fig:RSSconsumption} shows the memory
usage of the Bartender, the Librarian and one of the shepherds in the
last 37 hours of the test. Perhaps most noteworthy is that the memory
usage is very stable. The main reason for this is due to the way
Python, the programming language of Chelonia, allocates memory. As
memory allocation is an expensive procedure, Python tends to allocate
slightly more memory than needed and avoids releasing the already
acquired memory. As a result, the memory utilization gets evened out
after a period of time even though the usage of the system
varies. During the run-time of the test, files of different sizes were
periodically uploaded to and deleted from the system. The slight jump
in memory usage for the Bartender (green line) was during an
extraordinary upload of a set of large files. This jump was followed
by an increase in memory for the Librarian when the files were
starting to be replicated between the Shepherds, thus causing extra
requests to the Librarian.

Figure~\ref{fig:CPUconsumption} shows the CPU load for six of the
services. While the load on the A-Hashes, the Bartender and the
Librarian are all below 2.5\% of the CPU, Shepherd-1 (bottom left) and
Shepherd-8 (bottom right) use around 10\% and 20\%,
respectively. While the difference between the Shepherds is due to the
fact that Shepherd-1 was run on a server with twice the number of
CPU's as the server of Shepherd-8, the difference between the
Shepherds and the other services is due to the usage pattern during
the test. To confirm that the files stored on the storage node are
healthy, the Shepherd calculates a checksum for each file, first when
the file is received and later periodically. In periods where no new
files are uploaded, the Shepherds use almost no CPU as already stored
files don't need frequent checksum calculations. However, when files
are frequently uploaded, deleted and re-uploaded, as was the case
during the test, the number of checksum calculations, and hence the
CPU load, increases significantly. This can particularly be seen on
January 1 and 4 when a set of extra large files were uploaded, causing
spikes in the CPU load of the two Shepherds. Note also that the spikes
occurs at the same time for both Shepherd, as should be expected in a
load balanced system.

\section{Related Work}
\label{sec:relatedwork}
There are a number of grid and cloud storage solutions on the market,
focused on different storage needs. While direct performance
comparison with Chelonia is beyond the scope of this paper, some
similarities and differences between Chelonia and related storage
solutions are worth mentioning.

In the cloud storage family, Amazon Simple Storage Service
(S3)~\cite{amazons3} promises unlimited storage and high
availability. Amazon uses a two-level namespace as opposed to the
hierarchical namespace of Chelonia. In the security model of S3, users
have to implicitly trust S3 entirely, whereas in Chelonia users and
services need to trust a common independent third party Certificate
Authority. Additionally, S3 lacks fine-grained delegation and access
control lists are limited to 100 principals, limiting the usability
for larger scientific communities~\cite{amazons3forsciencegrids}.

While Chelonia is designed for geographically distributed users and
data storage, Hadoop \cite{hadoop} with its file system HDFS is
directed towards physically closely-grouped clusters. HDFS builds on
the master-slave architecture where a single NameNode works as a
master and is responsible for the metadata whereas DataNodes are used to
store the actual data. Though similar to Chelonia's metadata service,
the NameNode cannot be replicated and may become a bottleneck in the
system. Additionally, HDFS uses non-standard protocols for
communication and security while Chelonia uses standard protocols like
HTTP(S), GridFTP and X509.

When compared to typical grid distributed data management solutions,
the closest resemblance with Chelonia is the combination of the
storage element Disk Pool Manager (DPM) and the file catalog
LCG~\footnote{LHC (Large Hadron Collider) Computing Grid} File
Catalog (LFC)~\cite{dpmlfc}. By registering all files uploaded to
different DPM's in LFC one can achieve a single uniform namespace
similar to the namespace of Chelonia. However, where Chelonia has a
strong coupling between the Bartenders, Librarians and Shepherds to
maintain a consistent namespace, DPM and LFC have no coupling such
that registration and replication of files is handled on the client
side. If a file is removed or altered in DPM, this may not be
reflected in LFC. In Chelonia, a change of a file has to be registered
through the Bartender and propagated to the Librarian before it is
uploaded to the Shepherd.

dCache~\cite{dCache} differs from Chelonia in that dCache has a
centralized set of core services while Chelonia is distributed by
design. dCache is a service-oriented storage system which combines
heterogeneous storage elements to collect several hundreds of
terabytes in a single namespace. Originally designed to work on a
local area network, dCache has proven to be useful also in a grid
environment, with the Nordic Data Grid Facility (NDGF) dCache
installation~\cite{DSSWithdCache} as the largest example. There, the
core components, such as the metadata catalogue, indexing service and
protocol doors are run in a centralized manner, while the storage
pools are distributed. Chelonia, designed to have multiple instances
of all services running in a grid environment, will not need a
centralized set of core services. Additionally, dCache is relatively
difficult to deploy and integrate with new applications. Being a more
light-weight and flexible storage solution, Chelonia aims more towards
new, less demanding, user groups which are generally less familiar
with grid solutions.

Scalla \cite{Scalla} differs from Chelonia in that Scalla is designed
for use on centralized clusters, while Chelonia is designed for a
distributed environment. Scalla is a widely used software suite
consisting of an xrootd server for data access and an olbd server for
building scalable xrootd clusters. Originally developed for use with
the physics analysis tool ROOT \cite{root} , xrootd offers data access
both through the specialized xroot protocol and through other
third-party protocols. The combination of the xrootd and olbd
components offers a cluster storage system designed for low latency,
high bandwidth environments. In contrast, Chelonia is optimized for
reliability, consistency and scalability at some cost of latency and
is more suitable for the grid environment where wide area network
latency can be expected to be high.

Unlike Chelonia, iRODS \cite{irods} does not provide any storage
itself but is more an interface to other, third-party storage
systems. Based on the client-server model, iRODS provides a flexible
data grid management system. It allows uniform access to heterogeneous
storage resources over a wide area network. Its functionality, with a
uniform namespace for several Data Grid Managers and file systems, is
quite similar to the functionality offered by our gateway
module. However, iRODS uses a database system for maintaining the
attributes and states of data and operations. This is not needed with
Chelonia's gateway modules.

\section{Future Work}
\label{sec:futurework}
In addition to the continous process of improvements and
code-hardening (based on user feedback) there are plans to add some
new features.

The security of the current one-time URL based file transfers could be
improved by adding to the URL a signed hash of the IP and the DN
of the user. In this way the file transfer service could do
additional authorization, allowing the file transfer only for the same
user with the same IP.

Because of the highly modular architecture of both Chelonia and the
ARC HED hosting environment, the means of communication between the
services could be changed with a small effort. This would enable less
secure but more efficient protocols to replace HTTPS/SOAP when
Chelonia is deployed inside a firewall. This modularity also allows
additional interfaces to Chelonia to be implemented easily. For
example, an implementation of the WebDAV protocol would make the
system accesible to standard clients built into the mainstream
operating systems.

Another possible direction for enhancing the functionality of Chelonia
is to add handling of SQL databases. In addition to files, the system
could store database objects and use databases as storage nodes to
store them. SQL databases allow running extensive queries to get the
desired information. In a distributed environment, high availability
and consistency is often ensured by the replication of data. Access to
multiple copies of the data in the system also allows queries to be
run in parallel. Consistent, multiple copies of the data also provdies
a simple, transparent platform for scalable access to the same data to
a large number of distributed clients.

\section{Conclusions}
\label{sec:conclusion}
Chelonia is a cloud-like storage solution with grid
capabilities. While its core services resembles those of a traditional
data grid, the single-entry interface and the capabilities resemble
those of storage clouds.

An important part of developing a distributed storage system is proper
testing of the system, both in terms of performance and stability. The
presented tests are designed to give an understanding of how Chelonia
behaves in a real-life environment while at the same time controlling
the environment enough to get interpretable results. The tests have
shown that Chelonia can handle both deep and wide hierarchies as
expected, both in LAN environments and in WAN environments. The system
has shown self-healing capabilities, both in terms of individual
service stops and in terms of file availability. Multiple clients have
accessed the system simultaneously with reasonable performance results
and, even more importantly, Chelonia has been heavily used for more
than a week with stable performance even with vital services being
shut down during the test.

\section{Acknowledgements}
\label{Acknowledgements}

We wish to thank Mattias Ellert for vital comments and
proof reading. Additionally, we would like to thank UPPMAX, NIIF and
USIT for providing resources for running the storage tests.

The work has been supported by the European Commission through the KnowARC
project (contract nr. 032691) and by the Nordunet3 programme through the NGIn
project.

\bibliography{arcstorage_refs}
\bibliographystyle{plain}

\end{document}